\documentclass{jpsj3}
\usepackage{bm}
\usepackage{color}
\title{%
Phase-Dependent Damping Rate of Josephson Plasmons in the Nonequilibrium
  Steady State
}
\author{%
  Takanobu Jujo\thanks{E-mail address: jujo@ms.naist.jp}
}
\inst{%
  Division of Materials Science,
  Graduate School of Science and Technology,
  Nara Institute of Science and Technology,
  Ikoma, Nara 630-0101, Japan
}
\abst{%
  We calculate the damping rate of Josephson plasmons due to thermally excited quasiparticles in a nonequilibrium steady state.
  It is known that the experimentally observed dependence of the damping rate on the phase difference behaves in the opposite way to that of the theory.
  It is shown that this problem can be solved by calculating the damping rate by incorporating changes in the electronic state at the resonant frequency.
  When the phase difference is large, the broadening of the one-particle state is large due to the presence of a dc current and brings about the effective thermalization, so that the nonequilibrium correction becomes small.
  Conversely, when the phase difference is small, the nonequilibrium correction becomes large and causes a smaller damping rate due to a reduction in the effective temperature.
  This dependence of the nonequilibrium correction on the phase difference results in a damping rate that is consistent with experimental results.
}
\begin{document}
\setlength{\textwidth}{504pt}
\setlength{\columnsep}{14pt}
\hoffset-23.5pt
\maketitle

  \section{Introduction}

  The Josephson current, which depends on the phase difference
  between two superconductors in the tunnel junction,
  has a non-dissipative current
  and a dissipative current,~\cite{josephson62,josephson65}
  and there is a so-called cosine term problem in the latter.
  The problem is that there is an inconsistency between
  theory and experiment about 
  the sign of the phase-dependent term of the
  dissipative current proportional to the cosine of 
  the phase difference.~\cite{barone,langenberg}
  Theory shows that the sign of this term is positive,~\cite{harris,poulsen}
  but experiments on the damping rate of the Josephson plasmon
  show that the sign of this term is
  negative.~\cite{pedersen72,soerensen,rudner}

  In recent years, this term has been studied in relation to
  the energy relaxation in superconducting qubits.~\cite{catelani,leppakangas}
  The experiment carried out at very low temperatures showed that in
  the limit of small excitation frequencies, the sign of the cosine term is
  positive,~\cite{pop} as in theory.
  As in this study
  experiments on superconducting qubits are mainly aimed at extremely low
  temperatures where thermally excited quasiparticles can be neglected.

  On the other hand, in previous experiments 
  in which the negative sign of the cosine term was observed,
  thermally excited quasiparticles are thought to be responsible
  for the damping rate of Josephson plasmons.
  This is because these experiments were conducted at relatively
  high temperatures ($T>0.5 T_c$ with
  $T$ and $T_c$ being the temperature and the transition temperature,
  respectively), and the theoretical
  calculations were also performed for the case where
  the thermal excitation gives a finite value of the damping rate.
  For this reason, the previous cosine term problem at
  relatively high temperatures, rather than at low temperatures
  where superconducting qubits work, is considered to be unsolved.

  Up until now, theoretical calculations for the cosine term problem
  have focused only on the sign of
  the cosine term.~\cite{hida,zorin,gulyan,yeyati}
  Experiments have shown that as the phase difference becomes smaller,
  the damping rate of Josephson plasmons decreases.
  Therefore, rather than extracting only the cosine term and
  examining its sign, it is necessary to examine the dependence
  of the entire term in the dissipative current on the phase difference.

  In this study, we consider that at the resonance frequency of Josephson
  plasmons, the one-particle state of electrons is in a nonequilibrium
  steady state due to the external field at this frequency.
  We calculate the damping rate based on this electronic state.
  A finite phase difference occurs under the dc current which changes
  the resonance frequency. 
  The presence of this phase difference causes a broadening of the peak
  around the gap edge in the density of state.
  Therefore the nonequilibrium correction differs depending
  on the magnitude of the phase difference, and
  the resulting variation of
  the effective temperature
  (this is known as the \'Eliashberg effect~\cite{eliashberg70})
  leads to the phase-dependent damping rate due to
  quasiparticle excitations.

  The structure of this paper is as follows.
  In Sect. 2, we give a formulation based on the quasiclassical approximation
  in the nonequilibrium steady state, and derive the expression
  of the damping rate from the current.
  Section 3 gives the results of the numerical calculations
  based on this formulation.
  First, we show the calculated results of the damping rate without
  correction terms, and show that it gives the same behavior
  as in the previous studies.
  Next, we give the results of the damping rate in a nonequilibrium
  steady state, and show that the nonequilibrium correction plays
  a major role in the dependence of the damping rate on the phase difference.
  Section 4 provides a discussion of these results.

\section{Formulation}

\subsection{The expression of the current}

We give the expression of the current from
which the damping rate is derived.
This quantity is experimentally
obtained from the impedance measurements,~\cite{pedersen72}
and the expression of the impedance is given as follows.
(We set $\hbar=c=1$ with $c$ the velocity of light.)
By introducing the capacitance $C=\epsilon S/4\pi d$
($d$ and $S$ are the thickness and the area of the junction,
respectively),
we obtain $Cd\omega^2 A_{\omega}+J_{\omega}+J^{ext}_{\omega}=0$
from the Maxwell equation,~\cite{dahm}
in which $J_{\omega}$ and $J^{ext}_{\omega}$
are the induced and external currents, respectively~\cite{pines}
($A_{\omega}$ is the vector potential
with its frequency $\omega$).
With use of
$J_{\omega}=
-CdA_{\omega}(\omega_{\phi}^2-i\omega\gamma_{\omega})$,
which is obtained below (Sect. 2.3),
the impedance ($Z_{\omega}$) is written as
$Z_{\omega}
=-i\omega dA_{\omega}/J_{\omega}^{ext}
=(i\omega /C)/(\omega^2
-\omega_{\phi}^2+i\omega\gamma_{\omega})$.
Here, $\omega_{\phi}^2=\omega_J^2{\rm cos}(\phi_L-\phi_R)$
with $\omega_J=\sqrt{2eI_0/C}$
and $I_0=(\pi\Delta/2eR_N){\rm tanh}
(\Delta/2T)$
being the
plasma frequency~\cite{anderson} and the critical current~\cite{ambegaokar}
in the Josephson junction,
respectively. 
$\Delta e^{i\phi_L}$ and
$\Delta e^{i\phi_R}$ are the superconducting order parameters
of two superconductors consisting the Josephson junction
(the magnitude of the gap $\Delta$ is
assumed to be equal for two superconductors),
and $R_N$ is the tunnel resistance in the normal state.
$\gamma_{\omega}$ is the damping rate and
its expression is given in Sect. 2.3.

The expression of the current $J_{\omega}$
is obtained by using
the quasiclassical Green functions.
The following relation
holds from the kinetic equations for quasiclassical
Green functions:~\cite{voutilainen}
\begin{equation}
  \begin{split}
  \int d\epsilon 
  {\rm Tr}[\hat{\tau}_3 \omega
    \hat{g}^{LK}_{\epsilon+\omega/2,\epsilon-\omega/2}]
    =&
\iint d\epsilon d\epsilon_1 {\rm Tr}[
    \hat{\tau}_3\hat{\Sigma}^{L+}_{\epsilon+\omega/2,\epsilon_1}
    \hat{g}^{LK}_{\epsilon_1,\epsilon-\omega/2}
    +    \hat{\tau}_3\hat{\Sigma}^{LK}_{\epsilon+\omega/2,\epsilon_1}
    \hat{g}^{L-}_{\epsilon_1,\epsilon-\omega/2}
\\&    -
        \hat{g}^{L+}_{\epsilon+\omega/2,\epsilon_1}
        \hat{\Sigma}^{LK}_{\epsilon_1,\epsilon-\omega/2}
        \hat{\tau}_3
            -
        \hat{g}^{LK}_{\epsilon+\omega/2,\epsilon_1}
        \hat{\Sigma}^{L-}_{\epsilon_1,\epsilon-\omega/2}
        \hat{\tau}_3].
  \end{split}
  \label{eq:current}
\end{equation}
$\hat{g}^{L+(-)}$ and $\hat{g}^{LK}$
are the retarded (advanced) and Keldysh quasiclassical
Green functions, respectively, and
$\hat{\Sigma}^{L+(-)}$ and $\hat{\Sigma}^{LK}$ are
the self-energies.
The superscripts $L$ and $R$ are indices representing two superconductors
consisting the Josephson junction.
$\hat{\tau}_3=(\begin{smallmatrix} 1&0\\0&-1
\end{smallmatrix})$, $\hat{\cdot}$ means
a matrix in the Nambu space,
and ${\rm Tr}(\cdot)$ indicates taking the trace.
This leads to the following relation between
the current ($J_t$) and the temporal variation of the charge ($Q_t$):
\begin{equation}
  \begin{split}
&  \frac{d Q_t^L}{d t}=
    -e
        \rho_0
    \int\frac{d\omega}{2\pi}
     \omega\int \frac{d\epsilon}{4}    
  {\rm Tr}[\hat{\tau}_3 
    \hat{g}^{LK}_{\epsilon+\omega/2,\epsilon-\omega/2}]
  e^{-i\omega t}
\\&   =
-e
\rho_0
\int\frac{d\omega}{2\pi}
       \iint \frac{d\epsilon d\epsilon_1}{4}   
   {\rm Tr}[
    \hat{\tau}_3\hat{\Sigma}^{L+}_{\epsilon+\omega/2,\epsilon_1}
    \hat{g}^{LK}_{\epsilon_1,\epsilon-\omega/2}
    +    \hat{\tau}_3\hat{\Sigma}^{LK}_{\epsilon+\omega/2,\epsilon_1}
    \hat{g}^{L-}_{\epsilon_1,\epsilon-\omega/2}
\\&    -
        \hat{g}^{L+}_{\epsilon+\omega/2,\epsilon_1}
        \hat{\Sigma}^{LK}_{\epsilon_1,\epsilon-\omega/2}
        \hat{\tau}_3
            -
        \hat{g}^{LK}_{\epsilon+\omega/2,\epsilon_1}
        \hat{\Sigma}^{L-}_{\epsilon_1,\epsilon-\omega/2}
        \hat{\tau}_3]
    e^{-i\omega t}
        =-J_t.    
      \label{eq:QIrelation}
  \end{split}
\end{equation}
($\rho_0$ is the density of states which is assumed to be
equal for two superconductors, and the volumes of
superconductors are set to be 1 and not written explicitly below.)

We take account of
the tunneling effect as the self-energy correction~\cite{volkov}
as follows.
\begin{equation}
  \hat{\Sigma}^{L\pm}_{\epsilon,\epsilon'}
  =
  \pi
    \rho_0
  \int\frac{d\epsilon_1 d\epsilon_2}{(2\pi)^2}
  \hat{t}^{LR}_{\epsilon-\epsilon_1}
  \hat{g}^{R\pm}_{\epsilon_1,\epsilon_2}
  \hat{t}^{RL}_{\epsilon_2-\epsilon'}.
\end{equation}
Here,
\begin{equation}
  \hat{t}^{LR}_{\omega}=
  \int dt
  e^{i\omega t}
  \begin{pmatrix}
    t'e^{-ied\tilde{A}_{t}} & 0 \\ 0 & -t'e^{ied\tilde{A}_{t}}\end{pmatrix}.
\end{equation}
($\tilde{A}_t=\int\frac{d\omega}{2\pi}e^{-i\omega t}A_{\omega}$.
The case of zero voltage is studied in this paper.)
$e^{\pm ied\tilde{A}_{t}}$ is replaced by $e^{\mp ied\tilde{A}_{t}}$ in
$\hat{t}^{RL}_{\omega}$.
With use of the perturbative expansion by $edA_{\omega}$
\begin{equation}
  \hat{\Sigma}^{L\pm}_{\epsilon,\epsilon'}
  \simeq
  \pi
  \rho_0
     {t'}^{2}
     \Bigl[
     \hat{\tau}_3\hat{g}^{R\pm}_{\epsilon,\epsilon'}\hat{\tau}_3
  +\int\frac{d\epsilon_1}{2\pi}
  ied\left(
    \hat{\tau}_3\hat{g}^{R\pm}_{\epsilon,\epsilon_1}
  A_{\epsilon_1-\epsilon'}
  -A_{\epsilon-\epsilon_1}
  \hat{g}^{R\pm}_{\epsilon_1,\epsilon'}\hat{\tau}_3
\right)
  -\int\frac{d\epsilon_1d\epsilon_1'}{(2\pi)^2}
  (ied)^2
    A_{\epsilon-\epsilon_1}
  \hat{g}^{R\pm}_{\epsilon_1,\epsilon_1'}
  A_{\epsilon_1'-\epsilon'}
  +\cdots
  \Bigr].
\label{eq:sigsim}
\end{equation}
We use the monochromatic field such as
\begin{equation}
  A_{\omega}=2\pi
  \bar{A}[\delta(\omega-\omega_0)+\delta(\omega+\omega_0)],
\end{equation}
and the oscillating terms ($\epsilon-\epsilon'=n\omega_0$
with $n\ne 0$)
will be neglected below:
\begin{equation}
  \hat{g}^{L+}_{\epsilon,\epsilon'}\simeq
  \hat{g}^{L+}_{\epsilon}2\pi\delta(\epsilon-\epsilon').
  \label{eq:staticg}  
\end{equation}
This approximation 
holds in the case that
the strength of the external field
is much smaller than $\omega_0$:~\cite{semenov}
$\pi\rho_0(t'ed\bar{A})^2\ll \omega_0$.
From Eqs. (\ref{eq:QIrelation})
and (\ref{eq:sigsim}) up to the linear response
the current $J_{t}$ 
is written as follows.
\begin{equation}
  \begin{split}
    &
    J_{t}=    
\frac{1}{4eR_N}
    \int \frac{d\omega}{2\pi}
\int \frac{d\epsilon}{4}\bigl\{
  {\rm Tr}[\hat{g}^{R+}_{\epsilon}\hat{\tau}_3\hat{g}^{LK}_{\epsilon-\omega/2}
+\hat{g}^{RK}_{\epsilon}\hat{\tau}_3\hat{g}^{L-}_{\epsilon-\omega/2}    
-\hat{g}^{L+}_{\epsilon+\omega/2}\hat{\tau}_3\hat{g}^{RK}_{\epsilon}
-\hat{g}^{LK}_{\epsilon+\omega/2}\hat{\tau}_3\hat{g}^{R-}_{\epsilon}]
2\pi  \delta(\omega)
  \\&
  -i{\rm Tr}[\hat{\tau}_3\hat{g}^{R+}_{\epsilon-\omega/2}
    \hat{\tau}_3\hat{g}^{LK}_{\epsilon-\omega/2}
    +
    \hat{\tau}_3\hat{g}^{RK}_{\epsilon-\omega/2}
    \hat{\tau}_3\hat{g}^{L-}_{\epsilon-\omega/2}  
    +
    \hat{\tau}_3\hat{g}^{L+}_{\epsilon+\omega/2}
  \hat{\tau}_3\hat{g}^{RK}_{\epsilon+\omega/2}
  +
  \hat{\tau}_3\hat{g}^{LK}_{\epsilon+\omega/2}
  \hat{\tau}_3\hat{g}^{R-}_{\epsilon+\omega/2}
  \\&
    -\hat{g}^{L+}_{\epsilon+\omega/2}\hat{g}^{RK}_{\epsilon-\omega/2}
    -\hat{g}^{LK}_{\epsilon+\omega/2}\hat{g}^{R-}_{\epsilon-\omega/2}
        -\hat{g}^{R+}_{\epsilon+\omega/2}\hat{g}^{LK}_{\epsilon-\omega/2}
    -\hat{g}^{RK}_{\epsilon+\omega/2}\hat{g}^{L-}_{\epsilon-\omega/2}
  ]edA_{\omega}\bigr\}
  e^{-i\omega t}
  \end{split}
  \label{eq:currentjrl}
\end{equation}
with $R_N=1/4\pi e^2\rho_0^2 {t'}^2$.
The first line of Eq. (\ref{eq:currentjrl}) reduces to
the supercurrent $I_0{\rm sin}(\phi_L-\phi_R)$.

\subsection{The nonequilibrium steady state}

In this subsection we derive the one-particle Green functions
which are used to calculate the current [Eq. (\ref{eq:currentjrl})].
These functions are modified by the tunneling
effect and the external field.
The latter effect makes the system in
the nonequilibrium state, and 
the kinetic equations for quasiclassical Green's functions
are written as follows.~\cite{eliashberg}
\begin{equation}
  \hat{\tau}_3(\epsilon\hat{\tau}_0-\hat{\Delta}_L)
  \hat{g}^{L\pm}_{\epsilon,\epsilon'}
  -\hat{g}^{L\pm}_{\epsilon,\epsilon'}
  (\epsilon'\hat{\tau}_0-\hat{\Delta}_L)\hat{\tau}_3
  -\int d\epsilon_1
  \left(
  \hat{\tau}_3\hat{\Sigma}^{L\pm}_{\epsilon,\epsilon_1}
  \hat{g}^{L\pm}_{\epsilon_1,\epsilon'}
  -\hat{g}^{L\pm}_{\epsilon,\epsilon_1}
  \hat{\Sigma}^{L\pm}_{\epsilon_1\epsilon'}\hat{\tau}_3
  \right)=0
  \label{eq:g+eq}
\end{equation}
and
\begin{equation}
  \begin{split}
&  \hat{\tau}_3(\epsilon\hat{\tau}_0-\hat{\Delta}_L)
  \hat{g}^{L(a)}_{\epsilon,\epsilon'}
  -\hat{g}^{L(a)}_{\epsilon,\epsilon'}
  (\epsilon'\hat{\tau}_0-\hat{\Delta}_L)\hat{\tau}_3
  -\int d\epsilon_1
  \left(
  \hat{\tau}_3\hat{\Sigma}^{L+}_{\epsilon,\epsilon_1}
  \hat{g}^{L(a)}_{\epsilon_1,\epsilon'}
  +  \hat{\tau}_3\hat{\Sigma}^{L(a)}_{\epsilon,\epsilon_1}
  \hat{g}^{L-}_{\epsilon_1,\epsilon'}
  -\hat{g}^{L+}_{\epsilon,\epsilon_1}
  \hat{\Sigma}^{L(a)}_{\epsilon_1\epsilon'}\hat{\tau}_3
  -\hat{g}^{L(a)}_{\epsilon,\epsilon_1}
  \hat{\Sigma}^{L-}_{\epsilon_1\epsilon'}\hat{\tau}_3  
  \right)
  \\&
  -\pi\rho_0\int d\epsilon_1 d\epsilon_2 d\epsilon_2'
  \Bigl\{
  \hat{\tau}_3\hat{t}^{LR}_{\epsilon-\epsilon_2}
  \left[
    (T^h_{\epsilon_2'}-T^h_{\epsilon_1})
    \hat{g}^{R+}_{\epsilon_2,\epsilon_2'}
    -(T^h_{\epsilon_2}-T^h_{\epsilon})
    \hat{g}^{R-}_{\epsilon_2,\epsilon_2'}
    \right]
  \hat{t}^{RL}_{\epsilon_2'-\epsilon_1}
  \hat{g}^{L-}_{\epsilon_1,\epsilon'}
\\&  -
    \hat{g}^{L+}_{\epsilon,\epsilon_1}\hat{t}^{LR}_{\epsilon_1-\epsilon_2}
  \left[
    (T^h_{\epsilon_2'}-T^h_{\epsilon'})
    \hat{g}^{R+}_{\epsilon_2,\epsilon_2'}
    -(T^h_{\epsilon_2}-T^h_{\epsilon_1})
    \hat{g}^{R-}_{\epsilon_2,\epsilon_2'}
    \right]
  \hat{t}^{RL}_{\epsilon_2'-\epsilon'}
  \hat{\tau}_3
  \Bigr\}
  =0
  \end{split}
    \label{eq:gaeq}
\end{equation}
with
$\hat{\tau}_0=\left(\begin{smallmatrix}1&0\\0&1\end{smallmatrix}\right)$,
$\hat{\Delta}_L=\left(
\begin{smallmatrix}
  0& \Delta^{i\phi_L}\\ \Delta^{-i\phi_L}&0
  \end{smallmatrix}\right)$, and
\begin{equation}
  T^h_{\epsilon}:={\rm tanh}\left(\frac{\epsilon}{2T}\right).
\end{equation}
    [The superscript (a) in $\hat{g}^{L(a)}$ means the anomalous
      term in Keldysh Green function.]
By omitting the oscillating part as noted above in
Eq. (\ref{eq:staticg}) and using
the perturbative expansion Eq. (\ref{eq:sigsim})
and
\begin{equation}
  \hat{g}^{LK}_{\epsilon}=
  T^h_{\epsilon}(\hat{g}^{L+}_{\epsilon}-\hat{g}^{L-}_{\epsilon})
  +\hat{g}^{L(a)}_{\epsilon},
\end{equation}
Eqs. (\ref{eq:g+eq}) and (\ref{eq:gaeq})
are rewritten as follows.
\begin{equation}
  \hat{\tau}_3(\epsilon\hat{\tau}_0-\hat{\Delta}_L
  -\hat{\Sigma}^{L\pm}_{\epsilon})
  \hat{g}^{L\pm}_{\epsilon}
  -\hat{g}^{L\pm}_{\epsilon}
  (\epsilon\hat{\tau}_0-\hat{\Delta}_L
  -\hat{\Sigma}^{L\pm}_{\epsilon})\hat{\tau}_3
  =0
  \label{eq:g+0eq}
\end{equation}
with
\begin{equation}
  \hat{\Sigma}^{L\pm}_{\epsilon}=\hat{\Sigma'}^{L\pm}_{\epsilon}
      +\pi\rho_0
  {t'}^2\hat{\tau}_3\hat{g}^{R\pm}_{\epsilon}\hat{\tau}_3
      +\pi\rho_0
    (t'ed\bar{A})^2\sum_{s=\pm 1}\hat{g}^{R\pm}_{\epsilon+s\omega_0},
        \label{eq:sigmalp}
\end{equation}
and
\begin{equation}
  \hat{\tau}_3(\epsilon\hat{\tau}_0-\hat{\Delta}_L
  -\hat{\Sigma}^{L+}_{\epsilon})
  \hat{g}^{L(a)}_{\epsilon}
  -\hat{g}^{L(a)}_{\epsilon}
  (\epsilon\hat{\tau}_0-\hat{\Delta}_L
  -\hat{\Sigma}^{L-}_{\epsilon})\hat{\tau}_3
  +\hat{\tau}_3 \hat{z}^{L(a)}_{\epsilon} \hat{g}^{L-}_{\epsilon}
  -\hat{g}^{L+}_{\epsilon} \hat{z}^{L(a)}_{\epsilon} \hat{\tau}_3
  =0
  \label{eq:ga0eq}
\end{equation}
with
\begin{equation}
\hat{z}^{L(a)}_{\epsilon}=
  -\hat{\Sigma}^{L(a)}_{\epsilon}
      -\pi\rho_0
  (t'ed\bar{A})^2\sum_{\pm}(T^h_{\epsilon\pm\omega_0}-T^h_{\epsilon})
    (\hat{g}^{R+}_{\epsilon\pm\omega_0}-\hat{g}^{R-}_{\epsilon\pm\omega_0})
\end{equation}
and
\begin{equation}
  \hat{\Sigma}^{L(a)}_{\epsilon}=\hat{\Sigma'}^{L(a)}_{\epsilon}
    +\pi\rho_0
  {t'}^2\hat{\tau}_3\hat{g}^{R(a)}_{\epsilon}\hat{\tau}_3
      +\pi\rho_0
    (t'ed\bar{A})^2\sum_{\pm}\hat{g}^{R(a)}_{\epsilon\pm\omega_0}.
    \label{eq:sigmala}
\end{equation}
Here, $\hat{\Sigma'}_{\epsilon}^{L\pm}$ and
$\hat{\Sigma'}_{\epsilon}^{L(a)}$ in the above equations
represent the inelastic scattering effect which
is needed to
dissipate absorbed energies and 
maintain the nonequilibrium steady state
under the external fields, and the expression
of this effect will be specified below.
There also exist the equations with $L$ and $R$ interchanged
in the above equations.

We introduce the following phase transformation
to describe the current-carrying state due to the finite
phase difference ($\phi:=\phi_L-\phi_R\ne 0$);
\begin{equation}
\hat{\varphi}=
  \begin{pmatrix}  e^{i(\phi_L+\phi_R)/4} & 0 \\ 0 & e^{-i(\phi_L+\phi_R)/4}  \end{pmatrix}
  \label{eq:phasetrans}
\end{equation}
and multiply 
the right (left) of
Eq. (\ref{eq:g+0eq})
by $\hat{\varphi}$ (its complex conjugate $\hat{\varphi}^*$).
This equation 
is rewritten as 
\begin{equation}
  \hat{\tau}_3(\tilde{\epsilon}^{\pm}\hat{\tau}_0
  -\tilde{\Delta}^{\pm}_{1,\epsilon}\hat{\tau}_1
+i\tilde{\Delta}^{\pm}_{2,\epsilon}\hat{\tau}_2
  )
  \hat{g}^{\pm}_{\epsilon}
  -\hat{g}^{\pm}_{\epsilon}
(\tilde{\epsilon}^{\pm}\hat{\tau}_0
  -\tilde{\Delta}^{\pm}_{1,\epsilon}\hat{\tau}_1
+i\tilde{\Delta}^{\pm}_{2,\epsilon}\hat{\tau}_2
)\hat{\tau}_3=0.
\label{eq:eqforg+}
\end{equation}
Here, we put
\begin{equation}
  \hat{g}^{\pm}_{\epsilon}
  =
  \begin{pmatrix}
    g^{\pm}_{\epsilon} & f^{\pm}_{\epsilon}-h^{\pm}_{\epsilon}\\
    f^{\pm}_{\epsilon}+h^{\pm}_{\epsilon} & g^{\pm}_{\epsilon}
  \end{pmatrix}
:=
    \hat{\varphi}^*  \hat{g}^{L\pm}_{\epsilon} \hat{\varphi},
  \label{eq:gvarphi}
\end{equation}
\begin{equation}
  \tilde{\epsilon}^{\pm}=
  \epsilon-{\Sigma'}^{\pm}_{n,\epsilon}
   -\pi\rho_0
  {t'}^2g^{\pm}_{\epsilon}
    -\pi\rho_0
  (t'ed\bar{A})^2\sum_{s=\pm 1}g^{\pm}_{\epsilon+s\omega_0},
\end{equation}
\begin{equation}
  \tilde{\Delta}^{\pm}_{1,\epsilon}=
  \Delta_1+{\Sigma'}^{\pm}_{a,\epsilon}
    -\pi\rho_0
  {t'}^2f^{\pm}_{\epsilon}
    +\pi\rho_0
  (t'ed\bar{A})^2\sum_{s=\pm 1}f^{\pm}_{\epsilon+s\omega_0},
\end{equation}
\begin{equation}
  \tilde{\Delta}^{\pm}_{2,\epsilon}=
  \Delta_2+{\Sigma'}^{\pm}_{b,\epsilon}
    +\pi\rho_0
  {t'}^2h^{\pm}_{\epsilon}
    -\pi\rho_0
  (t'ed\bar{A})^2\sum_{s=\pm 1}h^{\pm}_{\epsilon+s\omega_0}
\end{equation}
with $\Delta_1=\Delta{\rm cos}(\phi/2)$,
$\Delta_2=-i\Delta{\rm sin}(\phi/2)$,
$\hat{\tau}_1=\left(\begin{smallmatrix} 0&1\\1&0\end{smallmatrix}\right)$,
$\hat{\tau}_2=\left(\begin{smallmatrix} 0&-i\\i&0\end{smallmatrix}\right)$,
and
\begin{equation}
{\Sigma'}^{\pm}_{n,\epsilon}\hat{\tau}_0      
+{\Sigma'}^{\pm}_{a,\epsilon}\hat{\tau}_1
-i{\Sigma'}^{\pm}_{b,\epsilon}\hat{\tau}_2=
\hat{\varphi}^*\hat{\Sigma'}^{L\pm}_{\epsilon}\hat{\varphi}.
\label{eq:sigmavarphi}
\end{equation}
Then
the solutions to Eq. (\ref{eq:eqforg+}) are
written as
\begin{equation}
  g^{\pm}_{\epsilon}=
  \frac{-\tilde{\epsilon}^{\pm}}{\tilde{\zeta}^{\pm}_{\epsilon}},
  \quad
  f^{\pm}_{\epsilon}=
  \frac{-\tilde{\Delta}^{\pm}_{1,\epsilon}}
       {\tilde{\zeta}^{\pm}_{\epsilon}},
       \quad
\text{and}\quad
       h^{\pm}_{\epsilon}=
         \frac{-\tilde{\Delta}^{\pm}_{2,\epsilon}}
              {\tilde{\zeta}^{\pm}_{\epsilon}}
              \label{eq:tildegfh}
\end{equation}
with $\tilde{\zeta}^{\pm}_{\epsilon}=
\sqrt{(\tilde{\Delta}^{\pm}_{1,\epsilon})^2
  -(\tilde{\Delta}^{\pm}_{2,\epsilon})^2
  -(\tilde{\epsilon}^{\pm})^2}$.
From $\hat{\varphi}^*\hat{g}^{R\pm}_{\epsilon}\hat{\varphi}$
and $\hat{\varphi}^*\hat{\Sigma'}^{R\pm}_{\epsilon}\hat{\varphi}$,
only the sign of $\hat{\tau}_2$ component is reversed,
and other than that, the same results as the above equations
are obtained.

Similarly Eq. (\ref{eq:ga0eq}) is rewritten as
\begin{equation}
  \begin{split}
&  \hat{\tau}_3(\tilde{\epsilon}^+\hat{\tau}_0
  -\tilde{\Delta}^+_{1,\epsilon}\hat{\tau}_1
+i\tilde{\Delta}^+_{2,\epsilon}\hat{\tau}_2
  )
  \hat{g}^{(a)}_{\epsilon}
  -\hat{g}^{(a)}_{\epsilon}
(\tilde{\epsilon}^-\hat{\tau}_0
  -\tilde{\Delta}^-_{1,\epsilon}\hat{\tau}_1
+i\tilde{\Delta}^-_{2,\epsilon}\hat{\tau}_1
  )\hat{\tau}_3
\\&
+\hat{\tau}_3
  (z_{0,\epsilon}\hat{\tau}_0
  +z_{1,\epsilon}\hat{\tau}_1
-iz_{2,\epsilon}\hat{\tau}_2
  )
  \hat{g}^{-}_{\epsilon}
  -\hat{g}^{+}_{\epsilon}
  (z_{0,\epsilon}\hat{\tau}_0
  +z_{1,\epsilon}\hat{\tau}_1
-iz_{2,\epsilon}\hat{\tau}_2
  )
  \hat{\tau}_3
  =0.
  \end{split}
  \label{eq:eqforga}
\end{equation}
Here,
$\hat{g}^{(a)}_{\epsilon}=\hat{\varphi}^*
\hat{g}^{(a)L}_{\epsilon} \hat{\varphi}$,
\begin{equation}
  z_{0,\epsilon}=
  -{\Sigma'}^{(a)}_{n,\epsilon}
    -\pi\rho_0
  {t'}^2g^{(a)}_{\epsilon}
    -\pi\rho_0
  (t'ed\bar{A})^2\sum_{\pm}g^{(a)}_{\epsilon\pm\omega_0}
    -\pi\rho_0
  (t'ed\bar{A})^2\sum_{\pm}
  (T^h_{\epsilon\pm\omega_0}-T^h_{\epsilon})
  (g^+_{\epsilon\pm\omega_0}-g^-_{\epsilon\pm\omega_0}),
  \end{equation}
\begin{equation}
  z_{1,\epsilon}=
  -{\Sigma'}^{(a)}_{a,\epsilon}
    +    \pi\rho_0
  {t'}^2f^{(a)}_{\epsilon}
    -\pi\rho_0
  (t'ed\bar{A})^2\sum_{\pm}f^{(a)}_{\epsilon\pm\omega_0}
    -\pi\rho_0
  (t'ed\bar{A})^2\sum_{\pm}
  (T^h_{\epsilon\pm\omega_0}-T^h_{\epsilon})
    (f^+_{\epsilon\pm\omega_0}-f^-_{\epsilon\pm\omega_0}),
\end{equation}
and
\begin{equation}
  z_{2,\epsilon}=
  -{\Sigma'}^{(a)}_{b,\epsilon}
    -    \pi\rho_0
  {t'}^2 h^{(a)}_{\epsilon}
    +\pi\rho_0
  (t'ed\bar{A})^2\sum_{\pm}h^{(a)}_{\epsilon\pm\omega_0}
    +\pi\rho_0
  (t'ed\bar{A})^2\sum_{\pm}
  (T^h_{\epsilon\pm\omega_0}-T^h_{\epsilon})
    (h^+_{\epsilon\pm\omega_0}-h^-_{\epsilon\pm\omega_0}).
\end{equation}
[${\Sigma'}^{(a)}_{n,\epsilon}$,
${\Sigma'}^{(a)}_{a,\epsilon}$,
and ${\Sigma'}^{(a)}_{b,\epsilon}$ are
the components of $\hat{\varphi}^*\hat{\Sigma'}^{L(a)}_{\epsilon}
\hat{\varphi}$ as in Eq. (\ref{eq:sigmavarphi}).
The $\hat{\tau}_3$-component is omitted because
it is confirmed to be small by numerical calculations
and its contribution to the damping rate is negligible.]
By introducing the quantity
\begin{equation}
  x^{(a)}_{\epsilon}
  :=\frac{g^{(a)}_{\epsilon}}{g^+_{\epsilon}-g^-_{\epsilon}}
  =\frac{f^{(a)}_{\epsilon}}{f^+_{\epsilon}-f^-_{\epsilon}}
  =\frac{h^{(a)}_{\epsilon}}{h^+_{\epsilon}-h^-_{\epsilon}},
\end{equation}
in which the equality of these three quantities
follows from the normalization conditions
for the quasiclassical Green functions,
Eq. (\ref{eq:eqforga})
is reduced to
\begin{equation}
  \begin{split}
    &  x^{(a)}_{\epsilon}
    [
    (-{\Sigma'}^+_{n,\epsilon}+{\Sigma'}^-_{n,\epsilon})
    (g^+_{\epsilon}-g^-_{\epsilon})
    -
        ({\Sigma'}^+_{a,\epsilon}-{\Sigma'}^-_{a,\epsilon})
    (f^+_{\epsilon}-f^-_{\epsilon})
    +
        ({\Sigma'}^+_{b,\epsilon}-{\Sigma'}^-_{b,\epsilon})
    (h^+_{\epsilon}-h^-_{\epsilon})
    ]
    \\&
    +
    {\Sigma'}^{(a)}_{n,\epsilon}
    (g^+_{\epsilon}-g^-_{\epsilon})
  +
    {\Sigma'}^{(a)}_{a,\epsilon}
    (f^+_{\epsilon}-f^-_{\epsilon})
      -
    {\Sigma'}^{(a)}_{b,\epsilon}
    (h^+_{\epsilon}-h^-_{\epsilon})
    +
        \pi\rho_0
    (t'ed\bar{A})^2\sum_{\pm}
    (\tilde{T}^h_{\epsilon\pm\omega_0}-\tilde{T}^h_{\epsilon})
    \\&\times
[(g^+_{\epsilon}-g^-_{\epsilon})
  (g^+_{\epsilon\pm\omega_0}-g^-_{\epsilon\pm\omega_0})
  +(f^+_{\epsilon}-f^-_{\epsilon})
  (f^+_{\epsilon\pm\omega_0}-f^-_{\epsilon\pm\omega_0})
  -(h^+_{\epsilon}-h^-_{\epsilon})
  (h^+_{\epsilon\pm\omega_0}-h^-_{\epsilon\pm\omega_0})
]=0,
  \end{split}
    \label{eq:kineqforxa}
\end{equation}
here,
\begin{equation}
  \tilde{T}^h_{\epsilon}:=
  T^h_{\epsilon}+x^{(a)}_{\epsilon}
  \label{eq:defxa}
\end{equation}
and $x^{(a)}_{\epsilon}$ has a meaning of
the nonequilibrium correction to the distribution function.
$x^{(a)}_{\epsilon}$ is obtained by solving Eq. (\ref{eq:kineqforxa}).

\subsection{Derivation of the damping rate from the current}

From Eq. (\ref{eq:currentjrl}),
the rf current term is written as
\begin{equation}
  \begin{split}
        & J_{\omega}=
    \frac{-i}{4eR_N}
\int \frac{d\epsilon}{4}
  {\rm Tr}[\hat{\tau}_3\hat{g}^{R+}_{\epsilon-\omega/2}
    \hat{\tau}_3\hat{g}^{LK}_{\epsilon-\omega/2}
    +
    \hat{\tau}_3\hat{g}^{RK}_{\epsilon-\omega/2}
    \hat{\tau}_3\hat{g}^{L-}_{\epsilon-\omega/2}  
    +
    \hat{\tau}_3\hat{g}^{L+}_{\epsilon+\omega/2}
  \hat{\tau}_3\hat{g}^{RK}_{\epsilon+\omega/2}
  +
  \hat{\tau}_3\hat{g}^{LK}_{\epsilon+\omega/2}
  \hat{\tau}_3\hat{g}^{R-}_{\epsilon+\omega/2}
  \\&
    -\hat{g}^{L+}_{\epsilon+\omega/2}\hat{g}^{RK}_{\epsilon-\omega/2}
    -\hat{g}^{LK}_{\epsilon+\omega/2}\hat{g}^{R-}_{\epsilon-\omega/2}
        -\hat{g}^{R+}_{\epsilon+\omega/2}\hat{g}^{LK}_{\epsilon-\omega/2}
        -\hat{g}^{RK}_{\epsilon+\omega/2}\hat{g}^{L-}_{\epsilon-\omega/2}]
  edA_{\omega}
  .
  \end{split}
  \label{eq:jwcurrent}
\end{equation}
Using the quasiclassical Green functions in the previous subsection,
this current is rewritten as
$J_{\omega}
=-CdA_{\omega}(\tilde{\omega}_{\phi}^2-i\omega\gamma_{\omega})$ with
$\tilde{\omega}_{\phi}^2$ and $\gamma_{\omega}$ being given as follows.
\begin{equation}
  \tilde{\omega}_{\phi}^2=
    \frac{1}{R_N C}
  \int d\epsilon
  \tilde{T}^h_{\epsilon}
  \left[
    -{\rm Re}(g^+_{\epsilon}-g^+_{\epsilon-\omega}){\rm Im}g^+_{\epsilon}
    +{\rm Re}(f^+_{\epsilon}-f^+_{\epsilon-\omega}){\rm Im}f^+_{\epsilon}
    +{\rm Im}(h^+_{\epsilon}-h^+_{\epsilon-\omega}){\rm Re}h^+_{\epsilon}
    \right]
  \label{eq:omegaphi2}
\end{equation}
and
\begin{equation}
  \gamma_{\omega}=
    \frac{1}{R_N C}
      \int d\epsilon
      \frac{(\tilde{T}^h_{\epsilon+\omega/2}-\tilde{T}^h_{\epsilon-\omega/2})}
      {2\omega}
      ({\rm Im}g^+_{\epsilon+\omega/2}{\rm Im}g^+_{\epsilon-\omega/2}
      +{\rm Im}f^+_{\epsilon+\omega/2}{\rm Im}f^+_{\epsilon-\omega/2}
      -{\rm Re}h^+_{\epsilon+\omega/2}{\rm Re}h^+_{\epsilon-\omega/2}).
      \label{eq:gammaomega}
\end{equation}
We introduce the dimensionless damping rate $\Gamma_{\omega}$ such
as
\begin{equation}
  \gamma_{\omega}=
  \frac{1}{R_N C}
  \Gamma_{\omega}.
      \label{eq:gammaomega2}
\end{equation}
$\Gamma_{\omega}= 
\Gamma^g_{\omega}+\Gamma^f_{\omega}$
with
\begin{equation}
  \Gamma^g_{\omega}:=
      \int d\epsilon
      \frac{(\tilde{T}^h_{\epsilon+\omega/2}-\tilde{T}^h_{\epsilon-\omega/2})}
      {2\omega}
      {\rm Im}g^+_{\epsilon+\omega/2}{\rm Im}g^+_{\epsilon-\omega/2}
      \label{eq:gammaomega2g}
\end{equation}
and
\begin{equation}
  \Gamma^f_{\omega}:=
      \int d\epsilon
      \frac{(\tilde{T}^h_{\epsilon+\omega/2}-\tilde{T}^h_{\epsilon-\omega/2})}
      {2\omega}
      ({\rm Im}f^+_{\epsilon+\omega/2}{\rm Im}f^+_{\epsilon-\omega/2}
      -{\rm Re}h^+_{\epsilon+\omega/2}{\rm Re}h^+_{\epsilon-\omega/2})
            \label{eq:gammaomega2f}.
\end{equation}
We call $\Gamma^g_{\omega}$ and $\Gamma^f_{\omega}$
the quasiparticle term and the interference term,~\cite{langenberg}
respectively.
The finite values of these two terms are brought about
by the thermal excitation of quasiparticles.

\section{Results of the Numerical Calculations}

\subsection{The damping rate in the absence of correction terms}
  
In this subsection we show results of
the damping rate in which 
the corrections by the tunneling effect
and the nonequilibrium distribution
are not included.
In this case quasiclassical Green functions,
Eq. (\ref{eq:tildegfh}), are approximated as
follows.
\begin{equation}
  g^{\pm}_{\epsilon}=\frac{-\epsilon}{\zeta^{\pm}_{\epsilon}},
  \quad
  f^{\pm}_{\epsilon}=\frac{-\Delta{\rm cos}(\phi/2)}{\zeta^{\pm}_{\epsilon}},
  \quad
  \text{and}\quad
  h^{\pm}_{\epsilon}=\frac{i\Delta{\rm sin}(\phi/2)}{\zeta^{\pm}_{\epsilon}},
  \label{eq:baregfh}
\end{equation}
here
$\zeta^{\pm}_{\epsilon}=\sqrt{\Delta^2-(\epsilon\pm i0)^2}$.
When these quantities and $x^{(a)}_{\epsilon}=0$ are used,
Eq. (\ref{eq:gammaomega}) results in
\begin{equation}
  \gamma_{\omega}^0=
      \frac{1}{R_N C}
(\Gamma^{g0}_{\omega}+\Gamma^{f0}_{\omega})
      \label{eq:gammaomega0}
\end{equation}
with
\begin{equation}
  \Gamma^{g0}_{\omega}:=
      \int d\epsilon
              \frac{({T}^h_{\epsilon+\omega/2}-{T}^h_{\epsilon-\omega/2})
                |\epsilon+\omega/2||\epsilon-\omega/2|
\theta(|\epsilon+\omega/2|-\Delta)\theta(|\epsilon-\omega/2|-\Delta)
              }
           {2\omega\sqrt{(\epsilon+\omega/2)^2-\Delta^2}
             \sqrt{(\epsilon-\omega/2)^2-\Delta^2}}
           \label{eq:gammaomega0g0}
      \end{equation}
and
\begin{equation}
  \Gamma^{f0}_{\omega}:={\rm cos}\phi
      \int d\epsilon
              \frac{({T}^h_{\epsilon+\omega/2}-{T}^h_{\epsilon-\omega/2})
                \Delta^2
\theta(|\epsilon+\omega/2|-\Delta)\theta(|\epsilon-\omega/2|-\Delta)
              }
                   {                     {\rm sgn}(\epsilon+\omega/2)
                     {\rm sgn}(\epsilon-\omega/2)
                    2\omega
 \sqrt{(\epsilon+\omega/2)^2-\Delta^2}
 \sqrt{(\epsilon-\omega/2)^2-\Delta^2}},
                   \label{eq:gammaomega0f0}
\end{equation}
here, $\theta(\cdot)$ means a step function.
This result is similar to 
that obtained previously in
Refs. 5 and 26.

We calculate
$\Gamma_{\omega}^0=
\Gamma_{\omega}^{g0}+\Gamma_{\omega}^{f0}$
numerically.
The superconducting gap at absolute zero
is set to be the unit of energy $\Delta_0=1$.
\begin{figure}
  \includegraphics[width=9cm]{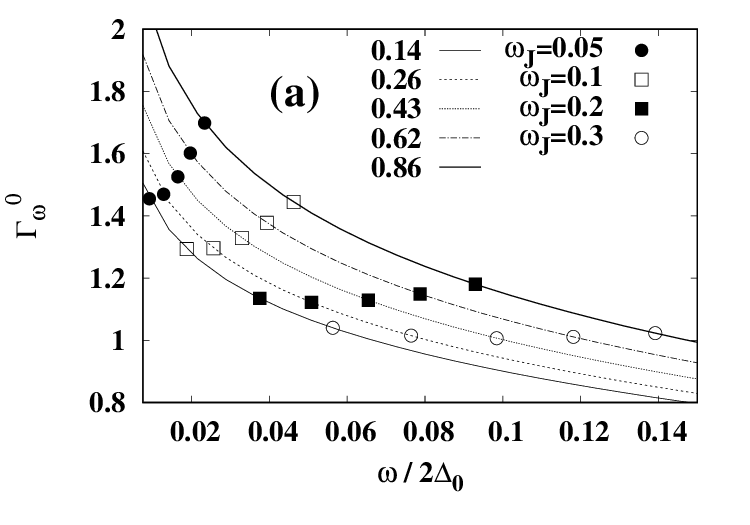}
  \includegraphics[width=9cm]{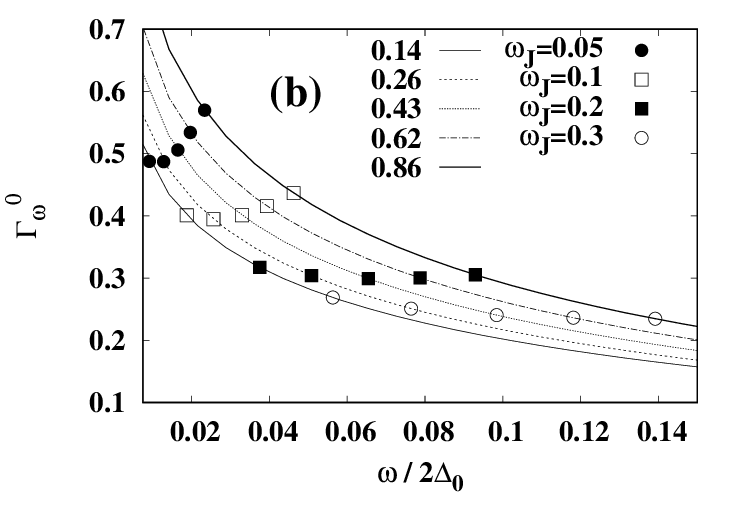}
  \caption{
    Dependence of the damping rate of Josephson plasmons
    on $\omega_J$ and ${\rm cos}\phi$ for calculations with use of
    the bare Green functions: $\Gamma_{\omega}^{0}=
    \Gamma_{\omega}^{g0}+\Gamma_{\omega}^{f0}$.
    The points indicate the damping rate 
    at the resonance frequency
    $\omega_0=\omega_J({\rm cos}\phi)^{1/2}$.
    The lines indicate $\Gamma_{\omega}^0$
    for several ${\rm cos}\phi$ whose values are indicated
    at the left of lines.
    For simplicity, $\omega_J/\Delta_0$ is written as
    $\omega_J$ in the figure.
    (a) $T/T_c=0.9$.
    (b) $T/T_c=0.5$.
      \label{fig:1}
    }
\end{figure}
Figure 1 shows the dependence of the damping rate $\Gamma_{\omega}^0$
on $\omega_J$ and ${\rm cos}\phi$.
The lines show the dependence of the damping rate on $\omega$
and each curve takes a different value of ${\rm cos}\phi$.
The points show the damping rate corresponding to each value of
$\omega_J$.
Since the possible range of values of $\omega_J({\rm cos}\phi)^{1/2}$
differs depending on the value of $\omega_J$,
the dependence of the damping rate on ${\rm cos}\phi$
also differs.
As the value of $\omega_J$ becomes smaller,
the tendency for the damping rate to increase becomes
stronger as a function of ${\rm cos}\phi$
because $\Gamma_{\omega}^0$ increases with decreasing $\omega$.
This causes the damping rate to have a positive slope
as a function of ${\rm cos}\phi$.
When the value of $\omega_J$ is large,
$\omega_J({\rm cos}\phi)^{1/2}$ takes a wide range of values,
so the dependence of the damping rate on ${\rm cos}\phi$
becomes small.
There is no significant difference between $T/T_c=0.9$
and $T/T_c=0.5$, but the positive slope as
a function of ${\rm cos}\phi$ is 
greater at higher temperatures.

\begin{figure}
  \includegraphics[width=9cm]{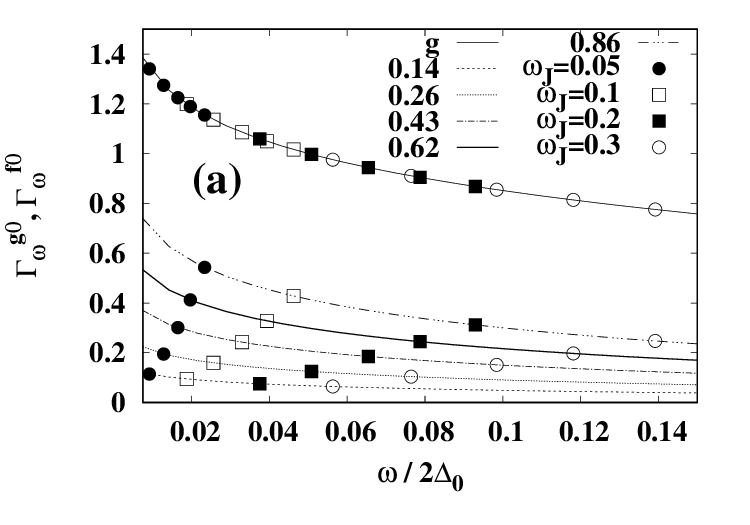}
  \includegraphics[width=9cm]{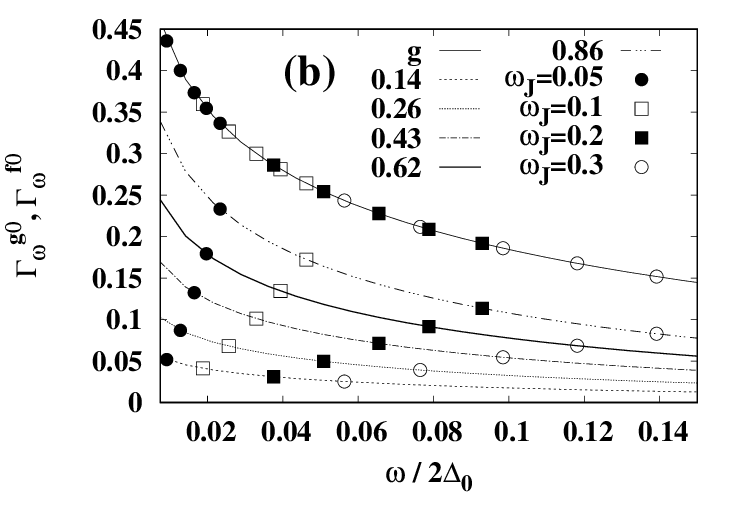}
  \caption{
    Dependence of $\Gamma_{\omega}^{g0}$ and $\Gamma_{\omega}^{f0}$
    on $\omega_J$ and ${\rm cos}\phi$.
    The points indicate the values at the resonance frequency
    $\omega_0=\omega_J({\rm cos}\phi)^{1/2}$.
    The upper line `g' indicates $\Gamma_{\omega}^{g0}$. 
    The five lines at the bottom
    indicate $\Gamma_{\omega}^{f0}$ for
    several ${\rm cos}\phi$ whose values are indicated
    at the left of the lines.
        For simplicity, $\omega_J/\Delta_0$ is written as
    $\omega_J$ in the figure.
    (a) $T/T_c=0.9$.
    (b) $T/T_c=0.5$.    
      \label{fig:2}
    }
\end{figure}

The above dependence of the damping rate on ${\rm cos}\phi$
mainly comes from the interference term $\Gamma_{\omega}^{f0}$
as shown in Fig.~\ref{fig:2}.
Figure 2 shows the decomposition of the damping rate 
into the quasiparticle term ($\Gamma_{\omega}^{g0}$)
and the interference term ($\Gamma_{\omega}^{f0}$).
As is clear from Eq. (\ref{eq:gammaomega0g0}),
the quasiparticle term has no explicit dependence on ${\rm cos}\phi$
in the absence of correction terms.
Then $\Gamma_{\omega}^{g0}$ takes values on the same curve
for all values of $\omega_J$ and ${\rm cos}\phi$,
and
the dependence of the quasiparticle term on ${\rm cos}\phi$
simply follows its dependence on $\omega$.
The damping rate due to the quasiparticle term becomes smaller
as the value of ${\rm cos}\phi$ increases. 
On the other hand, since the interference term is itself proportional
to ${\rm cos}\phi$ [Eq. (\ref{eq:gammaomega0f0})],
the dependence of $\Gamma_{\omega}^0$ on ${\rm cos}\phi$
is determined mainly by the interference term
rather than the quasiparticle term.
Then the damping rate $\Gamma_{\omega}^0$
increases with increasing the value of ${\rm cos}\phi$.
At higher temperatures, there are more thermally excited
quasiparticles, and the dependence of the damping rate
on $\omega$ becomes weaker.
The factor of ${\rm cos}\phi$
becomes more effective in the damping rate
at high $T$, and
this leads to the difference
between $T/T_c=0.9$ and $0.5$ in Fig. 1.

\subsection{The damping rate in the nonequilibrium steady state}

In this subsection, we consider the damping rate
when a nonequilibrium distribution function
is introduced as in Eq. ({\ref{eq:defxa})
  and the one-particle state is modified
  by a phase-dependent current as in Eq. (\ref{eq:tildegfh}).
Since the thermalization is
brought about by low-energy excitations,
we take account of the interaction between
electrons and acoustic phonons as the
self-energy due to inelastic scattering
introduced in Eqs. (\ref{eq:sigmalp}) and (\ref{eq:sigmala}).
The concrete expressions of
$\hat{\varphi}^*\hat{\Sigma'}^{L+}_{\epsilon}\hat{\varphi}$
and
$\hat{\varphi}^*\hat{\Sigma'}^{L(a)}_{\epsilon}\hat{\varphi}$
are similar to those of Eqs. (59) and (60) in
Ref. 27
except that $\hat{g}^{\pm}_{\epsilon}$ 
has $\hat{\tau}_2$-components in this calculation
as in Eq. (\ref{eq:gvarphi}).

Numerical calculations below are performed in the same way
as in
Ref. 27,
but there is a difference because of
the existence of the phase-dependent term.
Using Eq. (\ref{eq:omegaphi2}) 
with
$1/R_NC=\omega_J^2/\pi\Delta{\rm tanh}(\Delta/2T)$,
the numerically calculated value of the resonance frequency
$\sqrt{\tilde{\omega}_{\phi}^2}$
at $\omega=\omega_J({\rm cos}\phi)^{1/2}$
shows the same dependence on $\omega_J$ and ${\rm cos}\phi$
as $\omega_J({\rm cos}\phi)^{1/2}$ though
its absolute value is slightly shifted to
the high-frequency side under the influence of the
external field as in
Ref. 28.
The smallness of this shift is related to
that in the superconducting gap due to
the nonequilibrium correction, which
is small as compared to the change in the absorption
spectrum at frequencies below the gap edge as shown in
Fig. 4 of
Ref. 27.
We neglect this small shift, and use
$\sqrt{\tilde{\omega}_{\phi}^2}\simeq \omega_{\phi}=
\omega_J({\rm cos}\phi)^{1/2}$ as the resonance frequency
in the calculation of the damping rate below.

\begin{figure}
  \includegraphics[width=9cm]{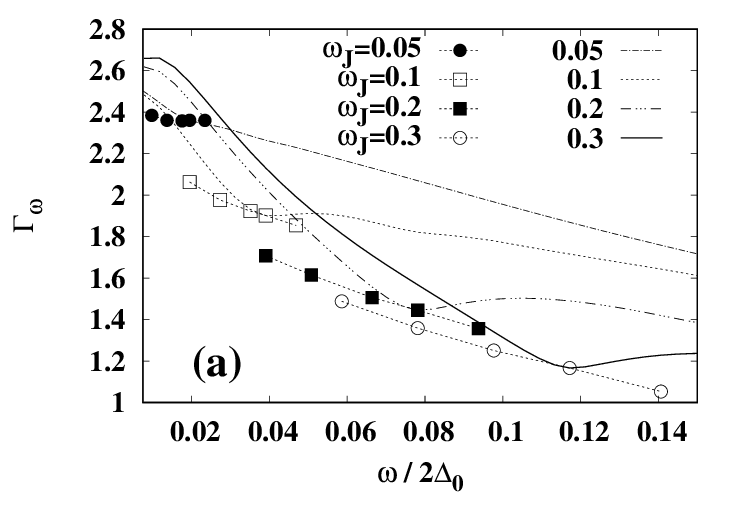}
  \includegraphics[width=9cm]{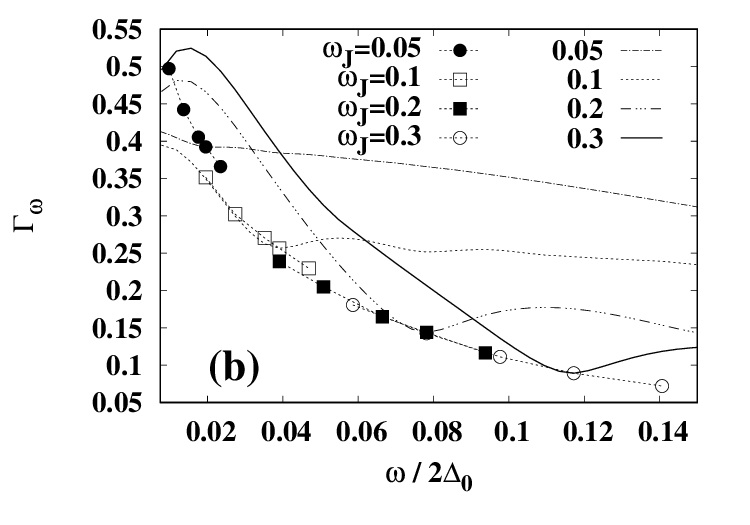}
  \caption{
    Dependence of the damping rate of Josephson plasmons 
    on $\omega_J$ and ${\rm cos}\phi$.
    The points indicate $\Gamma_{\omega_0}$ at the
    resonance frequency
    $\omega_0=\omega_J({\rm cos}\phi)^{1/2}$.
    The values of ${\rm cos}\phi$ are
    ``$0.15$, $0.30$, $0.49$, $0.61$, and $0.88$'' for
    $\omega_J/\Delta_0=0.05$ and $0.1$,
    ``$0.15$, $0.26$, $0.44$, $0.61$, and $0.88$'' for $\omega_J/\Delta_0=0.2$,
    and
    ``$0.15$, $0.27$, $0.42$, $0.61$, and $0.88$'' for $\omega_J/\Delta_0=0.3$.
          The four lines with no points indicate the dependence of
    $\Gamma_{\omega}$ on $\omega$ with
   ${\rm cos}{\phi}=0.61$ for $\omega_J/\Delta_0=0.05$,
   $0.1$, $0.2$, and $0.3$.
    $\pi\rho_0(t'ed\bar{A})^2=0.0005\Delta_0$, $\pi\rho_0t'^2=0.01\Delta_0$, and 
    $p'=0.05\Delta_0$.
    For simplicity, $\omega_J/\Delta_0$ is written as
    $\omega_J$ in the figure.
    (a) $T/T_c=0.9$.
    (b) $T/T_c=0.5$.
      \label{fig:3}
    }
\end{figure}
Figure 3 shows the damping rate of Josephson plasmons
$\Gamma_{\omega}$ at
several values of the resonance frequency
$\omega_J({\rm cos}\phi)^{1/2}$
in the nonequilibrium steady state.
[$p'$ is the coupling constant between electrons and
  acoustic phonons introduced
  in Eqs. (59) and (60) in
  Ref. 27.
      The value of $v_sk_F$ ($v_s$ is the velocity of sound)
  in these two equations is set to be $v_sk_F=8.0\Delta_0$
  in the numerical calculations as in Ref. 27, and
  this value will not be written explicitly hereafter.]
Unlike the case without these corrections in Fig. 1,
the dependence of the damping rate on $\omega$
differs depending on the value of $\omega_J$
because the nonequilibrium correction varies
depending on the resonance frequency.
The graphs of $\Gamma_{\omega}$
in the case of ${\rm cos}\phi=0.61$ are shown in Fig. 3,
but unlike Fig. 1, the 
dependence on $\omega$
differs depending on the values of $\omega_J$
due to the nonequilibrium correction.
(To avoid complexity, only the case of ${\rm cos}\phi=0.61$
is shown.)
It can be seen that the damping rate becomes smaller
with increasing ${\rm cos}\phi$, and
the sign of the slope of $\Gamma_{\omega_0}$
is negative as a function of ${\rm cos}\phi$ in
contrast to the case of $\Gamma_{\omega_0}^0$ in Fig. 1.
This is because, as shown below,
for the larger values of ${\rm cos}\phi$
the nonequilibrium correction becomes larger,
which results in a decrease in the damping rate.
This result of the dependence on ${\rm cos}\phi$
is different from previous theories~\cite{harris,poulsen}
on the damping rate,
and agrees with experimental results.~\cite{pedersen72,soerensen,rudner}
  The results for $T/T_c=0.9$ and $0.5$
  are similar except for the magnitude of the values,
  but there are differences in the damping rate
  as a function of ${\rm cos}\phi$,
  especially for small values of $\omega_J$.
At $T/T_c=0.9$ the dependence of $\Gamma_{\omega}$
on ${\rm cos}\phi$ is smaller than that of $T/T_c=0.5$.
This is related to the result that
the sign of the slope of $\Gamma_{\omega_0}$
as a function of ${\rm cos}\phi$
changes from negative to positive
as the temperature rises, which will be shown later. 
This also agrees with the experimental results
that the sign of the slope changes near $T_c$.

The effect of the nonequilibrium distribution on the damping rate
can be seen by considering the following quantities:
\begin{equation}
  \tau_x(\epsilon,\omega)=
  (x^{(a)}_{\epsilon+\omega/2}-x^{(a)}_{\epsilon-\omega/2})/\omega.
\end{equation}
The factor $(\tilde{T}^h_{\epsilon+\omega/2}-\tilde{T}^h_{\epsilon-\omega/2})
/\omega$ in Eq. (\ref{eq:gammaomega})
is equal to $\tau_h(\epsilon,\omega)+\tau_x(\epsilon,\omega)$
with $\tau_h(\epsilon,\omega)=
  (T^h_{\epsilon+\omega/2}-T^h_{\epsilon-\omega/2})/\omega$.
\begin{figure}
  \includegraphics[width=9cm]{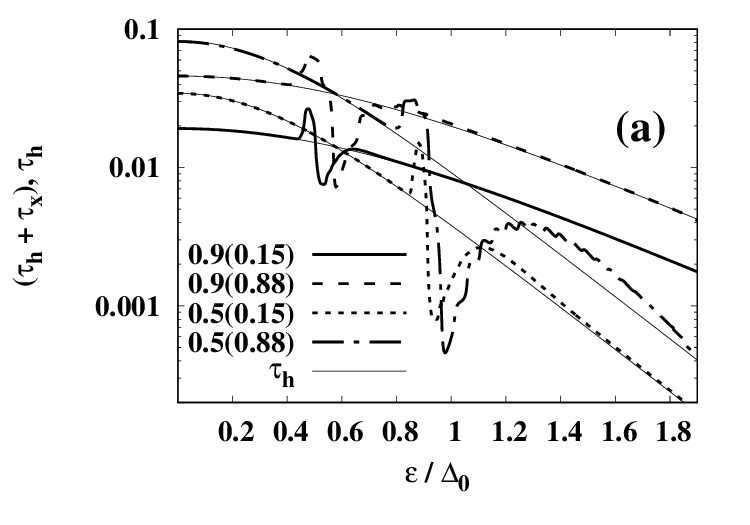}
  \includegraphics[width=9cm]{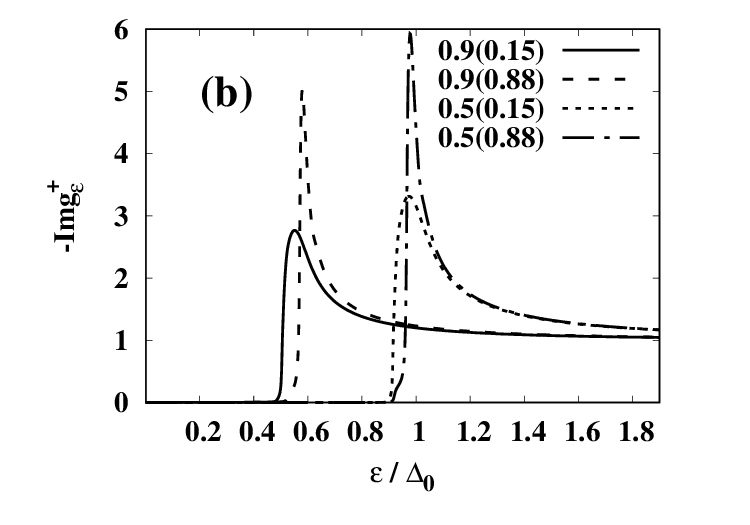}
  \caption{
    (a) Dependence of the nonequilibrium distribution function
    $\tau_h(\epsilon,\omega_0)+\tau_x(\epsilon,\omega_0)$
    on $\epsilon$
    at the resonance frequency
    $\omega_0=\omega_J({\rm cos}\phi)^{1/2}$
    at $T/T_c=0.9$ and $0.5$ with
    ${\rm cos}\phi=0.15$ and $0.88$.
    The four thin solid lines represent $\tau_h(\epsilon,\omega_0)$
    for the same values of $T/T_c$ and ${\rm cos}\phi$.
    (b) Dependence of the one-particle density of states
    on $\epsilon$
    in the presence of the supercurrent:
    $-{\rm Im}g^+_{\epsilon}$.
    The numbers on the left of lines
    indicate the values of $T/T_c$ (${\rm cos}\phi$).
    $\omega_J/\Delta_0=0.05$,
    $\pi\rho_0(t'ed\bar{A})^2=0.0005\Delta_0$, $\pi\rho_0t'^2=0.01\Delta_0$, and
    $p'=0.05\Delta_0$.
      \label{fig:4}
    }
\end{figure}
Figure 4(a) shows the effect of the nonequilibrium correction
$x^{(a)}_{\epsilon}$ on the thermal distribution ${\rm tanh}(\epsilon/2T)$.
For both $T/T_c=0.9$ and $0.5$,
the effect of $x^{(a)}_{\epsilon}$ becomes larger
with increasing ${\rm cos}\phi$.
Furthermore, the sign of $x^{(a)}_{\epsilon}$
changes near the energy gap and
becomes negative around energies at which
the density of states is large.
Therefore, $x^{(a)}_{\epsilon}$
has the effect of lowering the effective temperature.
Since the reduction in the effective temperature is greater at
the resonance frequencies with smaller phase differences,
the sign of the slope of the damping rate is negative
as a function of ${\rm cos}\phi$ as shown in Fig. 3.
This dependence of the nonequilibrium correction
on ${\rm cos}\phi$ is due to the
broadening of the density of states in the current-carrying
state.

Figure 4(b) shows the effect of the phase-dependent current
on the density of states. 
For both $T/T_c=0.9$ and $0.5$, 
the peak of the density of states becomes broad
with decreasing ${\rm cos}\phi$,
which is similar to the results in
Ref. 29.
This is because, as in the superconductors
including paramagnetic impurities,~\cite{abrikosov61,skalski}
the effect of time-reversal symmetry breaking
due to the phase-dependent current
on the density of states
becomes more pronounced
for the smaller values of ${\rm cos}\phi$.
However, as shown below, this variation of the density of states
itself
does not make a qualitative difference
in the dependence of the damping rate
on ${\rm cos}\phi$ as compared to the case without corrections.

\begin{figure}
  \includegraphics[width=9cm]{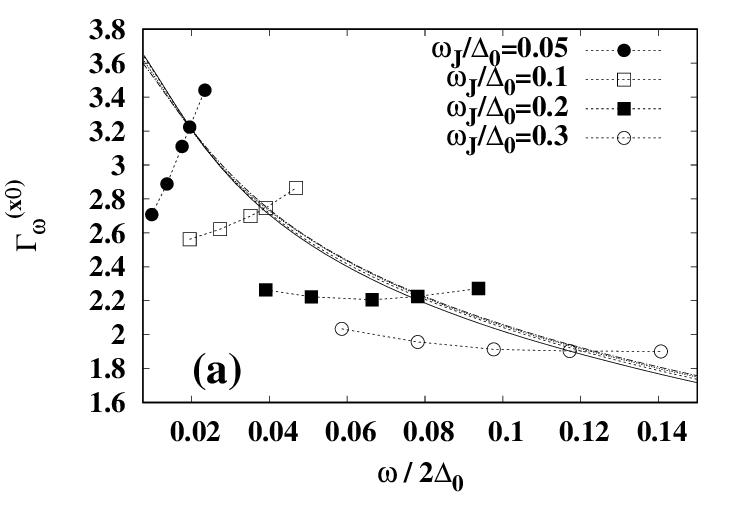}
  \includegraphics[width=9cm]{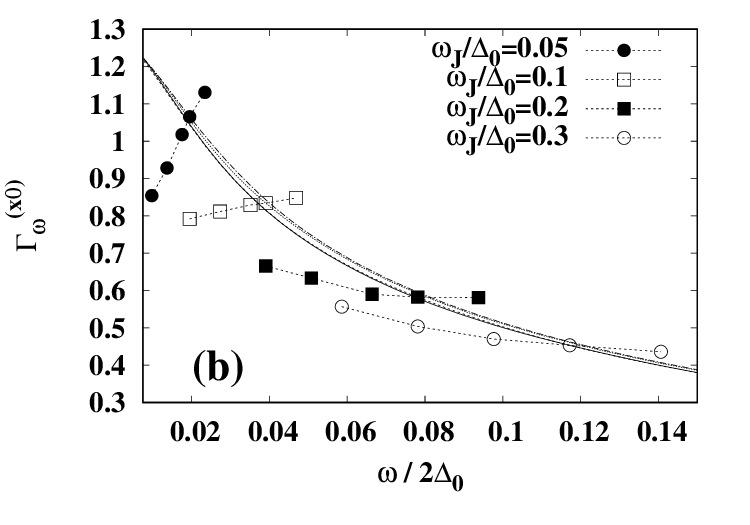}
  \caption{
    Dependence of the damping rate of Josephson plasmons on
    $\omega_J$ and ${\rm cos}\phi$
    in the absence of the nonequilibrium correction
    ($x^{(a)}=0$):
    $\Gamma_{\omega}^{(x0)}=\Gamma_{\omega}^{g(x0)}+\Gamma_{\omega}^{f(x0)}$.
    The points represent $\Gamma_{\omega_0}^{(x0)}$
    at the resonance frequency.
    The values of ${\rm cos}\phi$
    are the same as those in Fig. 3.
    The lines without points indicate the dependence of 
    $\Gamma_{\omega}^{(x0)}$ on $\omega$
    with ${\rm cos}\phi=0.61$
    for $\omega_J/\Delta_0=0.05$, $0.1$, $0.2$, and $0.3$.
    $\pi\rho_0(t'ed\bar{A})^2=0.0005\Delta_0$, $\pi\rho_0t'^2=0.01\Delta_0$, and 
    $p'=0.05\Delta_0$.
    (a) $T/T_c=0.9$.
    (b) $T/T_c=0.5$.
      \label{fig:5}
    }
\end{figure}
Figure 5 shows 
the damping rate at several values of
the resonance frequency $\omega_J({\rm cos}\phi)^{1/2}$
in the absence of the nonequilibrium correction:
$\Gamma_{\omega}^{(x0)}=\Gamma_{\omega}^{g(x0)}+\Gamma_{\omega}^{f(x0)}$.
Here, $\Gamma_{\omega}^{g(x0)}$ and $\Gamma_{\omega}^{f(x0)}$
are those of Eqs. (\ref{eq:gammaomega2g}) and (\ref{eq:gammaomega2f})
in which $x^{(a)}_{\epsilon}$ is set to be 0.
Unlike the results in Fig. 3,
the dependence of $\Gamma_{\omega}^{(x0)}$ on $\omega$ is hardly
affected by the value of $\omega_J$
and shows almost the same curve
when the value of ${\rm cos}\phi$ is the same
(the case of ${\rm cos}\phi=0.61$ is
shown for comparison).
This is almost the same as in the calculation
with the bare Green functions in Fig. 1.
This shows that the negative slope of the damping rate
$\Gamma_{\omega_0}$
as a function of ${\rm cos}\phi$ in Fig. 3 is caused
not by the current-induced change in the density of states
in Fig. 4(b),
but by the change in the nonequilibrium distribution
in Fig. 4(a).
By comparing the results between $T/T_c=0.9$ and $0.5$,
it can be seen that there is a difference
in the slope of the damping rate $\Gamma_{\omega_0}^{(x0)}$
which is similar to that in the case of Fig. 1.
The dependence of the damping rate on ${\rm cos}\phi$
arises from the interference term.
This can be seen by decomposing $\Gamma_{\omega_0}^{(x0)}$
into the quasiparticle term and the interference term.

\begin{figure}
  \includegraphics[width=9cm]{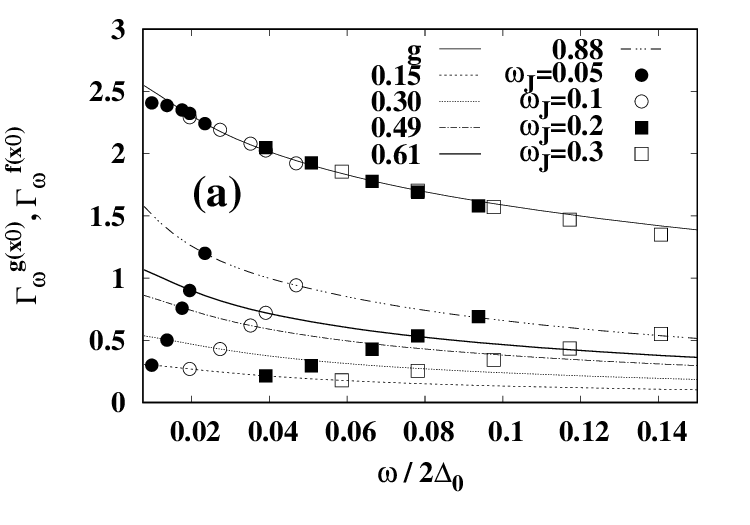}
  \includegraphics[width=9cm]{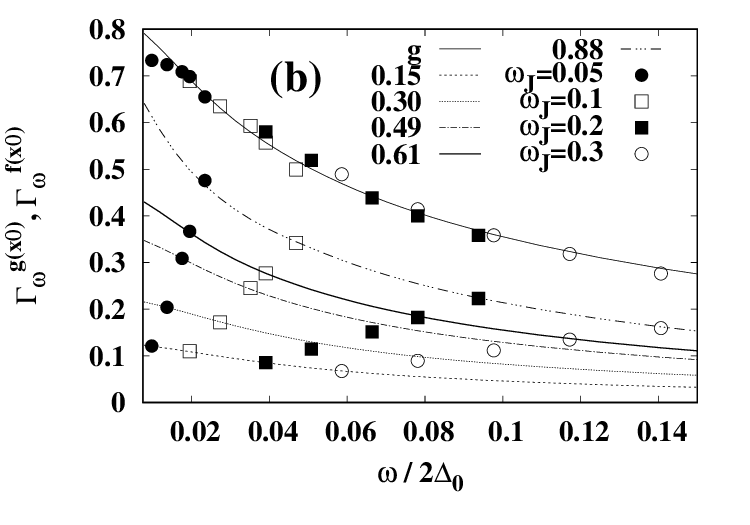}
  \caption{Dependence of the quasiparticle term
    ($\Gamma_{\omega}^{g(x0)}$)
    and the interference term ($\Gamma_{\omega}^{f(x0)}$) in
    the absence of the nonequilibrium correction ($x^{(a)}_{\epsilon}=0$)
    on $\omega_J$ and ${\rm cos}\phi$.
    The points indicate the values of these terms
    at the resonance frequency $\omega_0=\omega_J({\rm cos}\phi)^{1/2}$.
    The line `g' indicates the dependence of
    $\Gamma_{\omega}^{g(x0)}$ on $\omega$ with $\omega_J/\Delta_0=0.1$
    and ${\rm cos}\phi=0.61$.
    The five lines at the bottom show the
    dependence of $\Gamma_{\omega}^{f(x0)}$ on $\omega$
    with $\omega_J/\Delta_0=0.1$ and
    ${\rm cos}\phi$ (its value is shown to the left of lines).
    $\pi\rho_0(t'ed\bar{A})^2=0.0005\Delta_0$, $\pi\rho_0t'^2=0.01\Delta_0$, and
    $p'=0.05\Delta_0$.
            For simplicity, $\omega_J/\Delta_0$ is written as
    $\omega_J$ in the figure.
    (a) $T/T_c=0.9$.
    (b) $T/T_c=0.5$.    
      \label{fig:6}
    }
\end{figure}
Figure 6 shows the decomposition of the damping
rate $\Gamma_{\omega_0}^{(x0)}$
into the quasiparticle term
($\Gamma_{\omega}^{g(x0)}$)
and the interference term ($\Gamma_{\omega}^{f(x0)}$)
in the case of $x^{(a)}_{\epsilon}=0$.
The results are almost the same as those in Fig. 2,
and the quasiparticle term 
takes values on almost the same curve
regardless of the values of $\omega_J$ and ${\rm cos}\phi$.
(The reduction of the quasiparticle term at small
resonance frequencies is due to the broadening
of the density of states.)
The interference terms take values on different curves
depending on the values of ${\rm cos}\phi$,
but its dependence on $\omega$
hardly changes depending on the values of $\omega_J$.
This shows that, as mentioned above,
the dependence of the damping rate on ${\rm cos}\phi$
does not qualitatively change
in the presence of the phase-dependent current
without the nonequilibrium corrections.
There are studies in which the sign of the interference term
becomes negative when a phenomenological parameter
that smooths out the Riedel peak is introduced,~\cite{zorin,sheldon}
but this calculation showed that the sign of the interference term remains
positive in the current-carrying state.

When the nonequilibrium correction is taken into account,
the above behaviors of the quasiparticle and the interference
terms change qualitatively.
\begin{figure}
  \includegraphics[width=9cm]{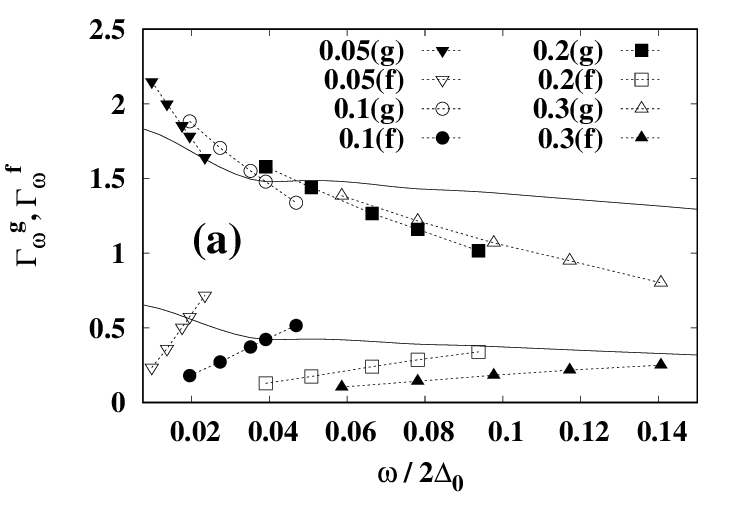}
  \includegraphics[width=9cm]{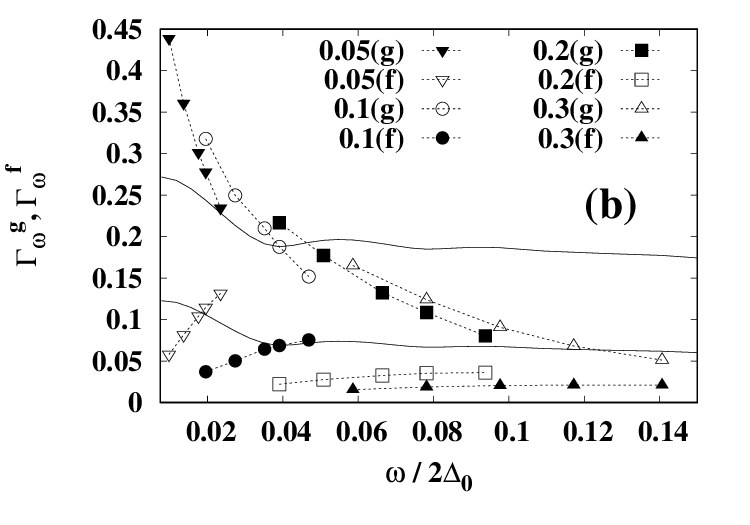}
  \caption{Dependence of the quasiparticle term ($\Gamma_{\omega}^{g}$)
    and the interference term ($\Gamma_{\omega}^{f}$)
    on $\omega_J$ and ${\rm cos}\phi$.
    The points represent the values of these terms 
    at the resonance frequency $\omega_0=\omega_J({\rm cos}\phi)^{1/2}$,
    and the numbers indicate the values of $\omega_J/\Delta_0$
    with $(g)$ and $(f)$ meaning 
$\Gamma_{\omega}^{g}$ and $\Gamma_{\omega}^{f}$, respectively.
    The values of ${\rm cos}\phi$ for each $\omega_J$ are
    the same as those in Fig. 3.
    The two solid lines show the dependence of
    $\Gamma_{\omega}^{g}$ (top)
    and $\Gamma_{\omega}^{f}$ (bottom) on $\omega$
    in the case of $\omega_J/\Delta_0=0.1$ and ${\rm cos}\phi=0.61$.
    $\pi\rho_0(t'ed\bar{A})^2=0.0005\Delta_0$, $\pi\rho_0t'^2=0.01\Delta_0$, and
    $p'=0.05\Delta_0$.
    (a) $T/T_c=0.9$.
    (b) $T/T_c=0.5$.    
      \label{fig:7}
    }
\end{figure}
Figure 7 shows the decomposition of the damping rate
$\Gamma_{\omega}$ in Fig. 3 into 
the quasiparticle term and the interference term
[Eqs. (\ref{eq:gammaomega2g}) and (\ref{eq:gammaomega2f}),
respectively].
Compared to the results in Figs. 2 and 6, the quasiparticle term has
a greater dependence on ${\rm cos}\phi$,
and 
the damping rate is suppressed 
with increasing ${\rm cos}\phi$ 
due to the effective temperature lowering
by the nonequilibrium correction.
In the same way
this correction suppresses also
the interference term
as a function of ${\rm cos}\phi$ as compared to the results
in Figs. 2 and 6.
These behaviors bring about
a negative slope in the damping rate $\Gamma_{\omega}$
as a function of ${\rm cos}\phi$
as shown in Fig. 3.

In Fig. 3, as a function of
${\rm cos}\phi$,  the negative slope at $T/T_c=0.9$
is less steep 
than that at $T/T_c=0.5$. 
This is attributed to the
large inelastic scattering effect at high temperatures,
which results in
relatively small nonequilibrium correction at high $T$
due to the strong thermalization
as shown in Fig. 4(a).
In Fig. 8, this effect of the inelastic scattering
is shown in a different way by
changing the values of the coupling constant ($p'$)
between electrons and acoustic phonons.
\begin{figure}
  \includegraphics[width=9cm]{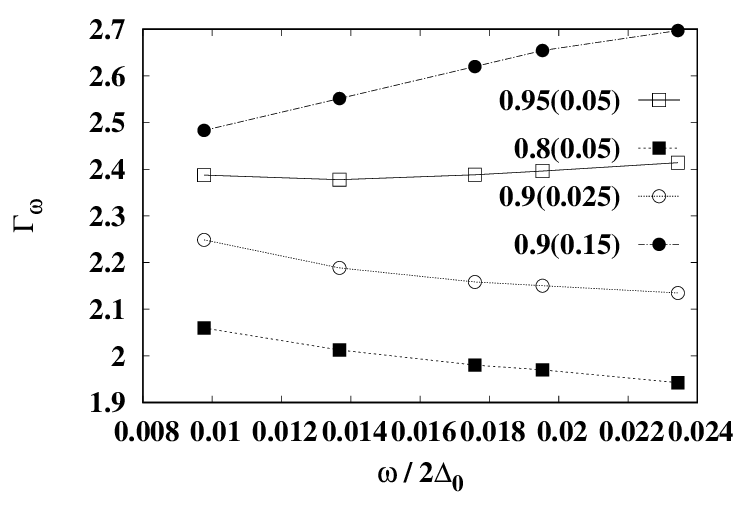}
  \caption{Dependence of the damping rate 
    of the Josephson plasmons on ${\rm cos}\phi$
    at the resonance frequency
    $\omega_0=\omega_J({\rm cos}\phi)^{1/2}$.
    The points indicate $\Gamma_{\omega_0}$ with 
    ${\rm cos}\phi=0.15$, $0.30$, $0.49$, $0.61$, and $0.88$.
    The numbers on the left of lines indicate
    the values of $T/T_c$ ($p'/\Delta_0$).
    $\omega_J/\Delta_0=0.05$, 
    $\pi\rho_0(t'ed\bar{A})^2=0.0005\Delta_0$, and $\pi\rho_0t'^2=0.01\Delta_0$,
      \label{fig:8}
    }
\end{figure}
Figure 8 shows the dependence of the damping rate
$\Gamma_{\omega}$ on ${\rm cos}\phi$ for
different values of $T/T_c$ and $p'$.
The sign of the slope of the damping rate as a function of
${\rm cos}\phi$ changes from negative to positive
as temperature increases for a fixed $p'$.
This change in sign occurs similarly when increasing $p'$
at a fixed temperature.
This shows that the inelastic scattering
induces a change in the sign of the slope
at high temperatures.

\section{Summary and Discussion}

In this paper, we calculated the dependence of the damping rate
on the phase difference at the resonance frequency.
We considered that the electronic state is in a nonequilibrium steady
state due to an external field with this resonance frequency.
In this state, the nonequilibrium distribution function is
dependent on the phase difference, and then the damping rate
changes depending on this difference.
When the phase difference is large, the time-reversal symmetry
breaking causes a broadening of the one-particle density of states
as in the presence of paramagnetic impurities.
As compared to the case where the phase difference is small
and the density of states has a sharp peak,
this broadening has effects of
strengthening the thermalization and reducing
the nonequilibrium correction.
Therefore, when the phase difference is small,
the reduction of the effective temperature
is larger than that in the case of a large phase difference.
This difference in the effective temperature depending on 
the phase difference causes the dependence of the damping
rate due to thermally excited quasiparticles
on the phase difference.

As a result, the damping rate becomes smaller with
decreasing phase difference (increasing ${\rm cos}\phi$).
On the other hand, the interference term has originally
an explicit dependence on the phase difference, which
causes an increase in the damping rate as the phase difference
becomes smaller.
Therefore, the actual dependence of the damping rate on the
phase difference is determined by the competition between
the nonequilibrium correction and the phase difference term
in the interference term.
The calculations in this paper showed that the former
can be predominant over the latter and the damping rate
becomes smaller with decreasing the phase difference.
This is consistent with experimental results showing that
the damping rate has a negative slope as a function of ${\rm cos}\phi$.
As the temperature rises, the effect of the inelastic scattering
by acoustic phonons increases, and the nonequilibrium correction
becomes smaller due to the strong thermalization.
Therefore, at high temperatures, 
the interference term becomes predominant in the damping rate,
and it is possible that the damping rate increases
with decreasing phase difference.
This means that the sign of the slope of the damping rate
as a function of ${\rm cos}\phi$ can change at high temperatures.
This behavior qualitatively agrees with the experimental results.

Other previous studies have assumed that the dependence of
the damping rate on the phase difference arises only from
the interference term, and have discussed the sign of
this term.~\cite{zorin,gulyan}
Since these studies considered the zero frequency limit,
the dependence of the quasiparticle term on the frequency
was not taken into account.
In experiments, the frequency is finite, and
$\omega_J/\Delta_0$ is about $1/20$.~\cite{pedersen72}
In our study, it was shown that even at a value of
$\omega_J/\Delta_0=0.05$, the dependence of the damping rate on the frequency
exists due to the nonequilibrium correction in the quasiparticle term.

Considering the comparison with the experiment,
in our calculation
$\gamma_{\omega}/\omega_J=
\Gamma_{\omega}/(R_NC\omega_J)\simeq$
$0.007$ and $0.038$ 
at $T/T_c=0.5$ and $T/T_c=0.9$, respectively,
by using an experimental value
$(1/R_NC)/\omega_J\simeq 0.016$.~\cite{pedersen72}
This is almost consistent with the experimental
value
$\gamma_{\omega}/\omega_J\simeq 0.014$
at $T/T_c=0.583$.
Though it is difficult to control the effect of inelastic scattering
experimentally, the magnitude of electron--phonon scattering
varies depending on materials.~\cite{kaplan}
Therefore, it is possible to verify this theory by 
examining the temperatures at which the dependence of the damping rate
on the phase difference changes 
for various materials.

\section*{Acknowledgment}

The numerical computation in this work was carried out
at the Yukawa Institute Computer Facility.

\end{document}